# UV Detector based on InAlN/GaN-on-Si HEMT Stack with Photo-to-Dark Current Ratio > $10^7$

Sandeep kumar[1a,b)], Anamika Singh Pratiyush[1a)], Surani B. Dolmanan[2], Sudhiranjan Tripathy[2], Rangarajan Muralidharan[1], Digbijoy N. Nath[1]

[1]Centre for Nano Science and Engineering (CeNSE),
Indian Institute of Science (IISc), Bangalore 560012, India
[2]Institute of Materials Research and Engineering (IMRE), Agency for Science, Technology, and Research (A*STAR), Innovis 08-03, 2 Fusionopolisway, Singapore 138634

**Abstract:**

**We demonstrate an InAlN/GaN-on-Si HEMT based UV detector with photo to dark current ratio > $10^7$. Ti/Al/Ni/Au metal stack was evaporated and rapid thermal annealed for Ohmic contacts to the 2D electron gas (2DEG) at the InAlN/GaN interface while the channel + barrier was recess etched to a depth of 20 nm to pinch-off the 2DEG between Source-Drain pads. Spectral responsivity (SR) of 34 A/W at 367 nm was measured at 5 V in conjunction with very high photo to dark current ratio of > $10^7$. The photo to dark current ratio at a fixed bias was found to be decreasing with increase in recess length of the PD. The fabricated devices were found to exhibit a UV-to-visible rejection ratio of >$10^3$ with a low dark current < 32 pA at 5 V. Transient measurements showed rise and fall times in the range of 3-4 ms. The gain mechanism was investigated and carrier lifetimes were estimated which matched well with those reported elsewhere.**

a) Sandeep Kumar and Anamika Singh Pratiyush have equally contributed for this work.
b) Corresponding author email: sandeepku@iisc.ac.in, sksandeep@cense.iisc.ernet.in

With applications in strategic sector, biomedical research, medical science, and UV astronomy, semiconductor ultra-violet photodetectors (PDs)attract widespread attention of the device and material research communities[1]. Compared to the conventional UV enhanced silicon photodiodes which suffer from poor efficiencies and necessitates the use of filters to reject the visible band for solar blind applications which make them bulky, inefficient and cumbersome, wide band gap materials, such as III-Nitrides[2–4] and $Ga_2O_3$[5–8] are intrinsically solar blind and offer high spectral responsivities. Different types of device geometries have been explored for GaN UV PDs (~365 nm) such as MSM[4,9], PN[3], PIN[10], and Schottky[2]. Besides these, III-nitride high electron mobility transistor (HEMT) stacks with a 2D electron gas (2DEG) – more commonly AlGaN/GaN – which constitute a matured transistor technology, are also being reported for UV detection at ~ 365 nm in their various configurations such as AlGaN/GaN gated[11], AlGaN/GaN 2DEG UV detectors[12], AlGaN/GaN with MESA in meandering geometry[13], GaN MSM with Al nanoparticles[14] and GaN MSM with Ag nanoparticles[15] have been reported earlier (listed in Table I). However, the spectral responsivity (SR) values of most of the PDs reported are comparatively low (< 5A/W at 1 V) while a few high SR values have also been reported in conjunction to high dark current (AlGaN/GaN 2DEG UV detectors). The high value of SR in meander geometry PDs[13] comes with a complex device design. This high SR value is also attributed to good material quality, as the HEMT stack was grown on SiC substrate. In this letter, we report a UV PD at 367 nm realized on a recess etched InAlN/GaN HEMT stack on silicon (111) with a high spectral responsivity of 7.5 A/W at 1 V. This device seeks to exploit the advantages of the highly conducting 2D electron gas in a III-nitride HEMT while ensuring that the dark current of such a design remains low by recess etching the 'gate' region between the source and drain.

The $In_{0.17}Al_{0.83}N$/GaN HEMT stack was used for fabricating UV PDs (schematic in Fig. 1(a)). The HEMT stack (total thickness ~1.9 μm) was grown on a Si (111) substrate using a metal organic chemical vapor deposition technique. The epilayers consist of ~10 nm $In_{0.17}Al_{0.83}N$ barrier, 1.0 nm AlN thin spacer, 150-nm undoped GaN channel, unintentionally doped GaN buffer, and AlGaN step-graded intermediate layers, overgrown on AlN/Si(111). The high-resolution reciprocal space mapping by x-ray diffraction confirmed an In concentration of ~17% in the HEMT barrier. The Hall data measured from these samples showed an average sheet resistance of 235 Ω/sq with a sheet carrier density of about $3.2 \times 10^{13}$ cm$^{-2}$. Atomic force microscopy measurements of such uncapped InAlN-based HEMT stack show an average surface rms roughness ~0.5 nm for 5.0 μm × 5.0 μm scan area. Following the standard lithographic process, Ti/Al/Ni/Au was evaporated for Ohmic contacts to $In_{0.17}Al_{0.83}N$/GaN stack. Ohmic contact annealing was performed at 850° C in $N_2$ ambient for 30 s. Mesa isolation of devices was done by dry etching using $BCl_3/Cl_2$ gas chemistry in reactive ion etching (RIE). The Ohmic metal pads (350 μm × 350 μm) of the fabricated devices were 300 μm spaced and the recess region was defined at the center, as shown in the Fig. 1 (b) between the Source and the Drain. Recess lengths of L = 3 μm, 5 μm, 7 μm, and 9 μm were opened using lithography for various devices, and 10 W of RF power in $BCl_3/Cl_2$ gas chemistry was used to recess etch to a depth of 20 nm, essentially etching the barrier + channel layer so as to completely deplete the 2D electron in this region. The SR measurement system used in this work is reported elsewhere[5].

Figure 2 shows SR versus wavelength (λ) characteristics at different bias voltages for a device with 3 μm recess length. The PD exhibits a peak SR value of 7.5 A/W and 55.2 A/W at 367 nm for 1 V and 15 V respectively. The UV-to-visible rejection ratio is defined as the ratio of peak SR value at 367 nm to the peak SR value at 420 nm which is > $10^3$ and indicative of visible-blind nature of the PD. The high SR value could be attributed to the high quality GaN channel of the HEMT stack which is expected to lead to a high minority carrier lifetime for photo-generated carriers. The high 2DEG density at InAlN/GaN interface results in $R_C$ = 0.26 ohm-mm and $R_{SH}$ = 262 ohm/□ from TLM measurements. Compared to Hall data, a small increase in the $R_{SH}$ value observed from the TLM measurement, which may be due to the thermal processing of Ohmic contacts. Hence there is a negligible voltage drop across the access regions and most of the voltage appears across the recessed region. In Fig. 3 the peak SR value first increases and then tend to saturate slowly at higher bias voltages. For a given bias, the SR decreases as the recess length increases. This indicates that the responsivity (or gain) in the devices is transit time limited.

Figure 4 shows the photo (367 nm) and dark current versus bias voltage for the 3 μm recess length device. The measured photo and dark currents were 1.2 mA and 58 pA respectively at 20 V. The photo-to-dark current ratio for device was > $10^7$, which is the highest for any type of III-nitride UV detector in this spectral range. Inset in Figure 4 shows the photo-to-dark current ratio versus recess length at a bias voltage of 20 V. The photo-to-dark current ratio (20 V) was found to be decreasing with increase in the recess length, which is attributed to the decrease in electric field with increase in recess length.

Figure 5 shows the transient response of the PD measured at 5 V. The light was chopped at 100 Hz using optical chopper and then focused on to the DUT after being guided through the optical assembly. The PD was biased using SR570 current amplifier and the voltage transient response of the PD was measured using an oscilloscope. The rise and fall times (10% - 90% of value) were 3.6 ms and 4.2 ms respectively for devices with 3 μm recess lengths. There was no significant difference in the rise and fall times for PDs with recess lengths of 5 μm, 7 μm, and 9 μm (not shown here).

To further reveal the photo response of the InAlN/GaN-based UV photodetector, the voltage dependent photocurrent (367 nm) of the PD was measured at different light intensities. Figure 6 (a), shows a series of photo I-V curves under increasing incident light intensity (from 4.9 to 11.5 mW/cm$^2$). It can be observed from Fig. 6 (a) that the photocurrent increases with increasing light intensity at a particular bias voltage. Figure 6 (b) shows the photocurrent of the detector versus light intensity at 20 V for different recess lengths and their linear fit. The photocurrent increases linearly with the incident excitation light intensity at the wavelength of 367 nm for different recess lengths. The linear increase of photocurrent with light intensity shows that the detector has negligible trap related gain.

Figure 7 shows gain of the PD versus bias voltage of different recess length devices (3 μm to 9 μm). Assuming the external quantum efficiency (η) to be 100%, the theoretical responsivity (R$_{Th}$) of UV photodetectors having a detection range of 367-368 nm can be calculated using the expression below (where G is the gain):

$$SR_{Ideal} = \frac{\eta q}{h\nu} \quad (1(a))$$

$$SR_{Measured} = \frac{G\eta q}{h\nu} \quad (1(b))$$

The ideal SR value for 367 nm comes out to be 0.29 A/W. This value of ideal responsivity is well surpassed even at a bias voltage of 1 V, which is a clear indication of gain in the devices. The high gain (>10$^2$ at 15 V) resulting in high responsivity of the devices is due to photoconductive gain as the metal contacts have a non-rectifying (Ohmic) nature unlike in the more commonly reported MSM or Schottky geometries [16]. The photoconductive gain of an intrinsic photoconductor with Ohmic contacts on both electrodes is given by[17].

$$G = \frac{\tau_s}{\tau_{tr}^e} \quad (2(a))$$

$$\tau_{tr}^e = \frac{L^2}{\mu_e * V} \quad (2(b))$$

where, $\tau_s$ is the excess carrier lifetime, $\tau_{tr}^e$ is the transit time, L is the recess length and V is the applied voltage across the recess length of the PD. Using equations (2(a)), (2(b)) and assuming the mobility of electron in GaN recess channel from the earlier reported value to be 100 cm$^2$/V-s[18], we have estimated the excess carrier lifetime values (~ 20 ns), which is in good agreement with the earlier reported values[19]. It can be observed from figure 7, the gain in devices with 3 μm recess length increases linearly till 5 V and then tend to saturate slowly at higher voltages whereas, gain in devices with recess length of 5 μm, 7 μm and 9 μm tend to saturate at lower voltages. The early gain saturation in devices with longer recess lengths (9 μm) compared to smaller recess lengths (3 μm) is attributed to higher channel resistance in the former, which limits the photocurrent[20]. Hence, we expect higher photocurrent and gain for smaller recess devices. As can be observed from figure 8, the Y-axis intercept (Gain ~ 0 V) for 3 μm device is higher than 9 μm device. This suggests that high gain and photocurrent can be expected for submicron recess devices. Fabrication of submicron recess length PDs are underway, which are expected to exhibit even higher responsivity (gain) as they would not suffer early saturation of gain.

In summary, we demonstrate record high photo-to-dark current ratio exceeding 10$^7$ for UV photodetectors based on InAlN/GaN-on-Si HEMT stack. High responsivity of 32.9 A/W at 5 V is obtained with UV-to-visible rejection ratio >10$^3$. We have also demonstrated low dark current 32 pA at 5 V and transient response rise and fall time of 3.6 ms and 4.2 ms respectively. Photocurrent dependence on light intensity is also studied for devices with different recess lengths. The high gain is explained by photoconductive gain mechanism. This demonstration of a state-of-art UV detector in a HEMT stack holds promise towards physical integration of UV devices with high power/RF transistors on the same substrate which could open up exciting avenues to explore new devices with added functionalities.

This work was funded by the Department of Science and Technology (DST) under its Water Technology Initiative (WTI), Grant No. DST 01519. This publication is an outcome of the Research and Development work undertaken in the Project under Ph.D. scheme of Media

Lab Asia. We would also like to thank Ministry of Electronics and Information Technology (MeiTY) for the financial support for the work. Authors would also like to acknowledge the National NanoFabrication Centre (NNFC) and Micro and Nano Characterization Facility (MNCF) at CeNSE, IISc for device fabrication and characterization.

**Figure and legends**:

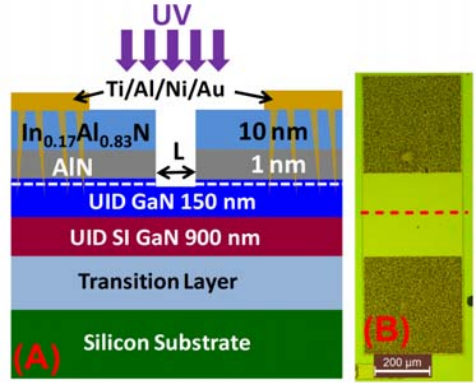

**Figure 1: (a)** Schematic of fabricated device from HEMT stack **(b)** Optical micrograph of the fabricated device. Recess region (3 μm) is shown by dashed line.

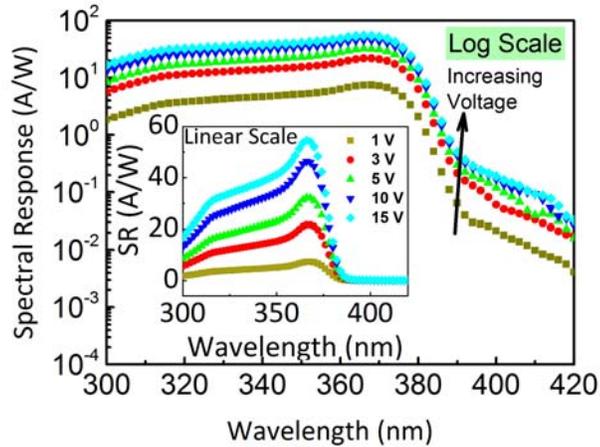

**Figure 2:** Spectral response versus wavelength (300 nm- 420 nm) of GaN PD (3 μm recess) (Log scale) at different bias voltages (1 V to 15 V), also showing UV to visible rejection ratio >$10^3$ at 5 V. The inset shows SR versus wavelength at different bias voltages (Linear Scale).

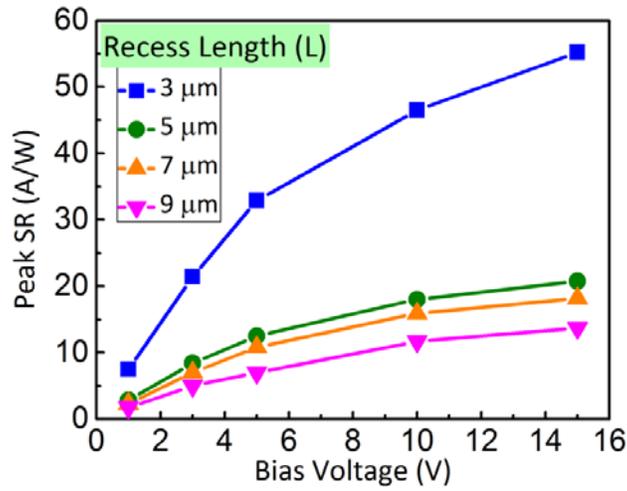

**Figure 3:** The peak SR versus bias voltage for different recess lengths (3 μm to 9 μm) PDs.

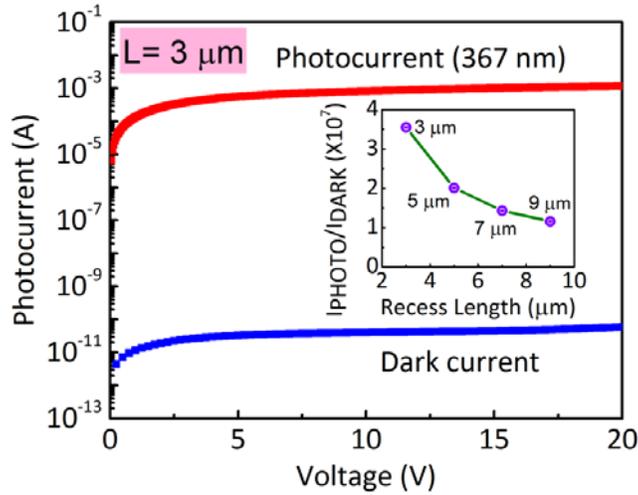

**Figure 4:** Photo (367 nm) and dark current for 3 um recess (Till 20 V). Inset shows photo-to-dark current ratio versus recess length at 20 V.

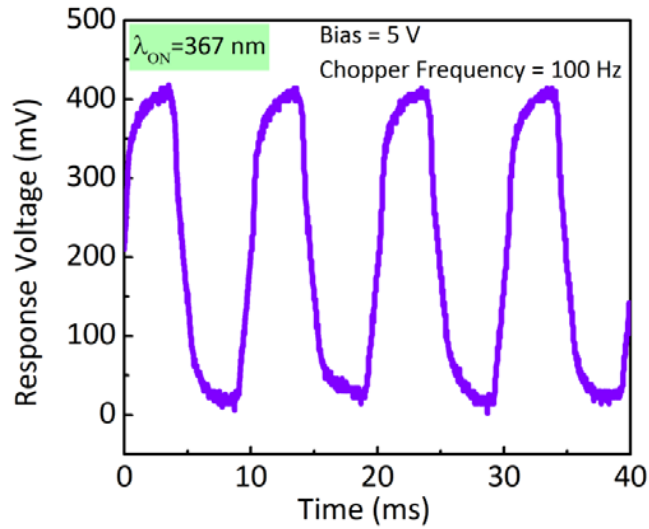

**Figure 5:** Transient response of the detector measured using CRO and SR570 current amplifier.

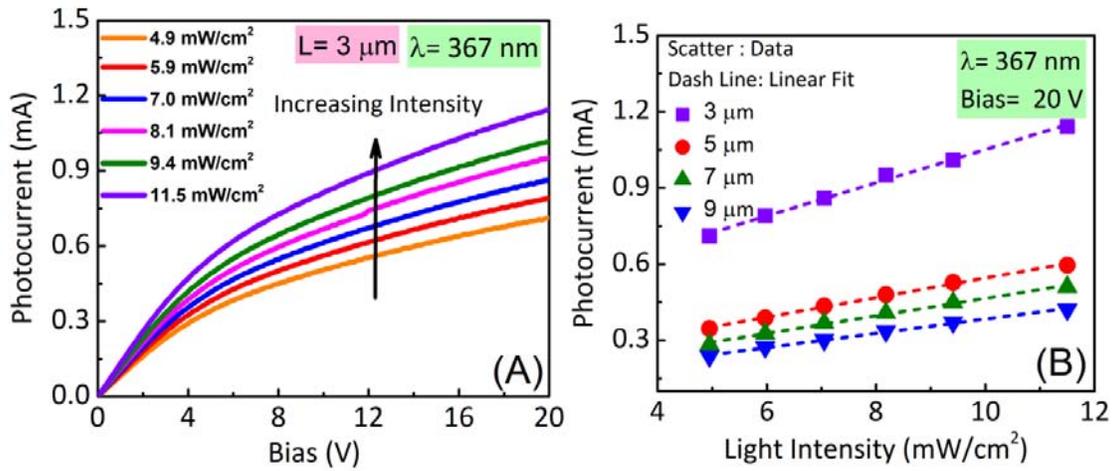

**Figure 6: (a)** Photocurrent (367 nm) versus voltage for 3 μm recess length device at different light intensities. **(b)** Photocurrent (367 nm) versus light intensity at 20 V for different recess length devices (3 μm to 9 μm). Dashed line shows liner fit to experimental data.

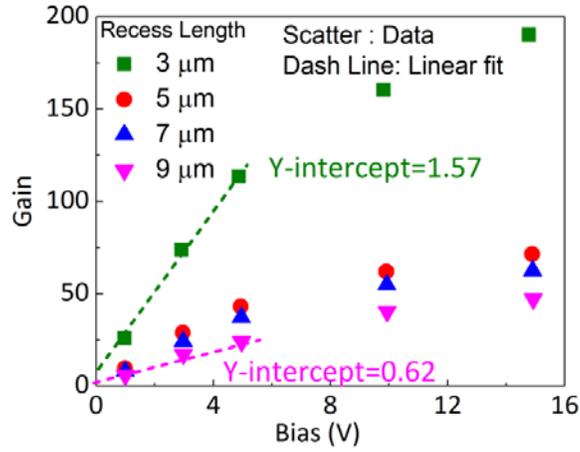

**Figure 7:** Gain versus bias voltage of different recess length devices (3 μm to 9 μm). Dashed line shows liner fit to experimental data up to 5 V.

**Tables**:

TABLE I. List of UV detectors (~365 nm, GaN band edge) reported earlier for different material systems.

| Material System | Detector Type | SR (Bias Voltage) | $I_{Photo}/I_{Dark}$ (Bias Voltage) | Reference |
|---|---|---|---|---|
| GaN | MSM | 0.5 A/W (5 V) | $10^7$ (5 V), $10^3$ (10 V) | 49 |
| GaN | PN | 0.15 A/W (-) | 50 (0.5 V) | 3 |
| GaN | PIN | 0.25 A/W (5 V) | $10^2$ (5 V) | 10 |
| GaN | Lateral Schottky | 0.02 A/W (0 V) | - | 2 |
| GaN | Vertical Schottky | 0.1 A/W (0 V) | - | 2 |
| AlGaN/GaN | Gated HEMT | 3 A/mW (10 V) | 1.25 (10 V) | 11 |
| AlGaN/GaN | HEMT 2DEG UV | $5.2 \times 10^9$ A/W (-) | 2 (6 V) | 12 |
| AlGaN/GaN | Meander MSM -Al | 10 A/mW (5 V) | $10^4$ (10 V) | 13 |
| GaN | Nanoparticle MSM- Ag | 3 A/W (10 V) | $10^6$ (10 V) | 14 |
| GaN | Nanoparticle | 0.14 A/W (5 V) | - | 15 |